\documentclass[useAMS,usenatbib]{mn2e}
\usepackage{times}
\usepackage{graphicx}
\usepackage{float}
\usepackage{amsmath}
\usepackage{amssymb}
\usepackage[bookmarks=true,bookmarksnumbered=true]{hyperref}
\title[Stellar Motion due to Gravitational Instabilites]{Stellar Motion Induced by Gravitational Instabilities in Protoplanetary Disks}
\author[S. Michael and R. H. Durisen]{Scott Michael$^{1}$\thanks{E-mail:
scamicha@indiana.edu}, R. H. Durisen$^{1}$\\
$^{1}$Department of Astronomy, Indiana University, Bloomington, IN 47405, USA}
\begin{document}

\pagerange{\pageref{firstpage}--\pageref{lastpage}} \pubyear{2009}

\maketitle

\label{firstpage}

\begin{abstract}

We test the effect of assumptions about stellar motion on the behavior of gravitational instabilities in protoplanetary disks around solar-type stars by performing two simulations that are identical in all respects except the treatment of the star. In one simulation, the star is assumed to remain fixed at the center of the inertial reference frame. In the other, stellar motion is handled properly by including an indirect potential in the hydrodynamic equations to model the star's reference frame as one which is accelerated by star/disk interactions. The disks in both simulations orbit a solar mass star, initially extend from 2.3 to 40 AU with a $\varpi^{-1/2}$ surface density profile, and have a total mass of 0.14 M$_{\odot}$. The $\gamma = 5/3$ ideal gas is assumed to cool everywhere with a constant cooling time of two outer rotation periods. 

The overall behavior of the disk evolution is similar, except for weakening in various measures of GI activity by about at most tens of percent for the indirect potential case. Overall conclusions about disk evolution in earlier papers by our group, where the star was always assumed to be fixed in an inertial frame, remain valid. There is no evidence for independent one-armed instabilities, like SLING, in either simulation. On the other hand, the stellar motion about the system center of mass (COM) in the simulation with the indirect potential is substantial, up to 0.25 AU during the burst phase, as GIs initiate, and averaging about 0.9 AU during the asymptotic phase, when the GIs reach an overall balance of heating and cooling. These motions appear to be a stellar response to nonlinear interactions between discrete global spiral modes in both the burst and asymptotic phases of the evolution, and the star's orbital motion about the COM reflects the orbit periods of disk material near the corotation radii of the dominant spiral waves. This motion is, in principle, large enough to be observable and could be confused with stellar wobble due to the presence of one or more super-Jupiter mass protoplanets orbiting at 10's AU. We discuss why the excursions in our simulation are so much larger than those seen in simulations by \citet{rice2003a}.

\end{abstract}


\section{Introduction}
\label{sec:intro}
As reviewed in \citet{durisen2007}, gravitational instabilities (GIs) will lead to the dynamic growth of spiral waves in protoplanetary disks whenever the Toomre parameter $Q = c_s\kappa/\pi G\Sigma$ becomes less than about 1.5 to 1.7 or so. Here, $c_s$ is the sound speed, $\kappa$ is the epicyclic frequency ($\approx \Omega$, the disk orbital angular frequency, in a nearly Keplerian disk), and $\Sigma$ is the  surface mass density. As the spirals grow to nonlinear amplitude, either they will saturate at a point where radiative cooling balances the heat generated by the waves, or they will fragment into bound clumps if the radiative cooling time $t_{cool}$ is comparable to or shorter than the disk rotation period $P_{rot} = 2\pi/\Omega$ \citep[e.g.,][]{gammie2001, rice2003b, rice2005, mejia2005, clarke2007}. In nonfragmenting disks, mass transport rates for realistic conditions can result in effective \citet{shakura1973} $\alpha$-viscosity parameters of order 10$^{-2}$ \citep[e.g.,][]{lodato2004, boley2006}, sufficient to dominate major phases of disk evolution \citep[e.g.,][]{vorobyov2007}. Fragmentation of disks due to fast cooling may be an important planet formation mechanism under some conditions \citep[e.g.,][]{boss1997, boss2000, boss2001, mayer2002, mayer2004, stamatellos2009, boley2009, clarke2009, rafikov2009}.

Early simulations of GIs in disks by the Indiana Univeristy hydrodynamics group (IUHG) included a hydrodynamically active star \citep{pickett1998, pickett2000}, but the disks we could model with active stars were only ten times larger than the stellar radius. To produce more realistically sized protoplanetary disks, we introduced a central hole and fixed the star at the center of our cylindrical computational grid \citep{pickett2003, mejia2005, boley2006, boley2007, caiphd2006, cai2008,boley2008}. The hydrodynamics equations remained written as if the grid reference frame were inertial. With an artificially fixed central stellar potential in the inertial frame, star/disk interactions that might shift the star off the system center of mass (COM) cannot be properly modeled. Acceleration of the star and subsequent feedback into the spiral structure is explicitly suppressed. In a sense, the star has infinite inertia and becomes a sink for momentum. Under these conditions, SLING amplification \citep{adams1989,shu1990} of spiral structure cannot be properly computed. The concern about our earlier calculations is then two-fold: 1) The physics in our simulations of GI-active disks may be inaccurate, e.g., leading to erroneous estimates of the effective $\alpha$s for mass transport, and may lack important features, e.g., stronger and more coherent one-armed spirals. 2.) The stellar motions that we suppress could be a significant and observable signature of GI activity in disks \citep{rice2003a}.

Authors have approached the problem of stellar motion for disk simulations in different ways. In Smoothed Particle Hydrodynamics (SPH) simulations of GIs \citep{rice2003b,mayer2004,lodato2004,stamatellos2008,forgan2009}, the stellar motion is included automatically by treating the star as a central sink particle that is smoothed differently from the rest of the SPH particles. However, these simulations, as in \citet{rice2003a}, can  exhibit a large initial accretion rate onto the central object. How the transfer of angular momentum from the disk material to the central object is handled could be quite important to the star's motion. The only global 3D grid-based hydrodynamics scheme other than the IUHG code so far used for published simulations of GIs in protoplanetary disks around solar-type stars is the Eulerian spherical-grid code of Alan Boss. Although he does not explicitly integrate the stellar equation of motion, \citet{boss2000} allows the central protostar to wobble in response to the growth of nonaxisymmetry in the disk by repositioning the star to keep the system COM at the grid center. There is no guarantee with this scheme that the star's displacement is a true response to Newtonian reaction forces applied by the disk. The star's motion could simply represent an accumulation of numerical error in the location of the COM. In principle, then, this treatment is no better than what was done in the IUHG simulations. In order to explore the effect of the star/disk interaction, we have now implemented the indirect potential method \citep[e.g.,][]{nelson2000b}. This effectively puts our simulations into the accelerated reference frame of the star and therefore properly accounts for stellar motion while keeping the star at the center of our computational grid. 

In this article, we present the results of a simulation for a nonfragmenting disk using the indirect potential to treat stellar motion and compare it to results of an identical simulation with an artificially fixed central star from Mej\'ia et al. (2005). In \S\ref{sec:results}, we first present an overall qualitative comparison, followed by a detailed analysis of the stellar motion and of differences in the disk behaviors. We compare our results to those obtained via other numerical methods and briefly discuss possible observational consequences in \S\ref{sec:compare}. Finally, \S\ref{sec:conclude} presents our main conclusions. 

\begin{figure*}
\includegraphics{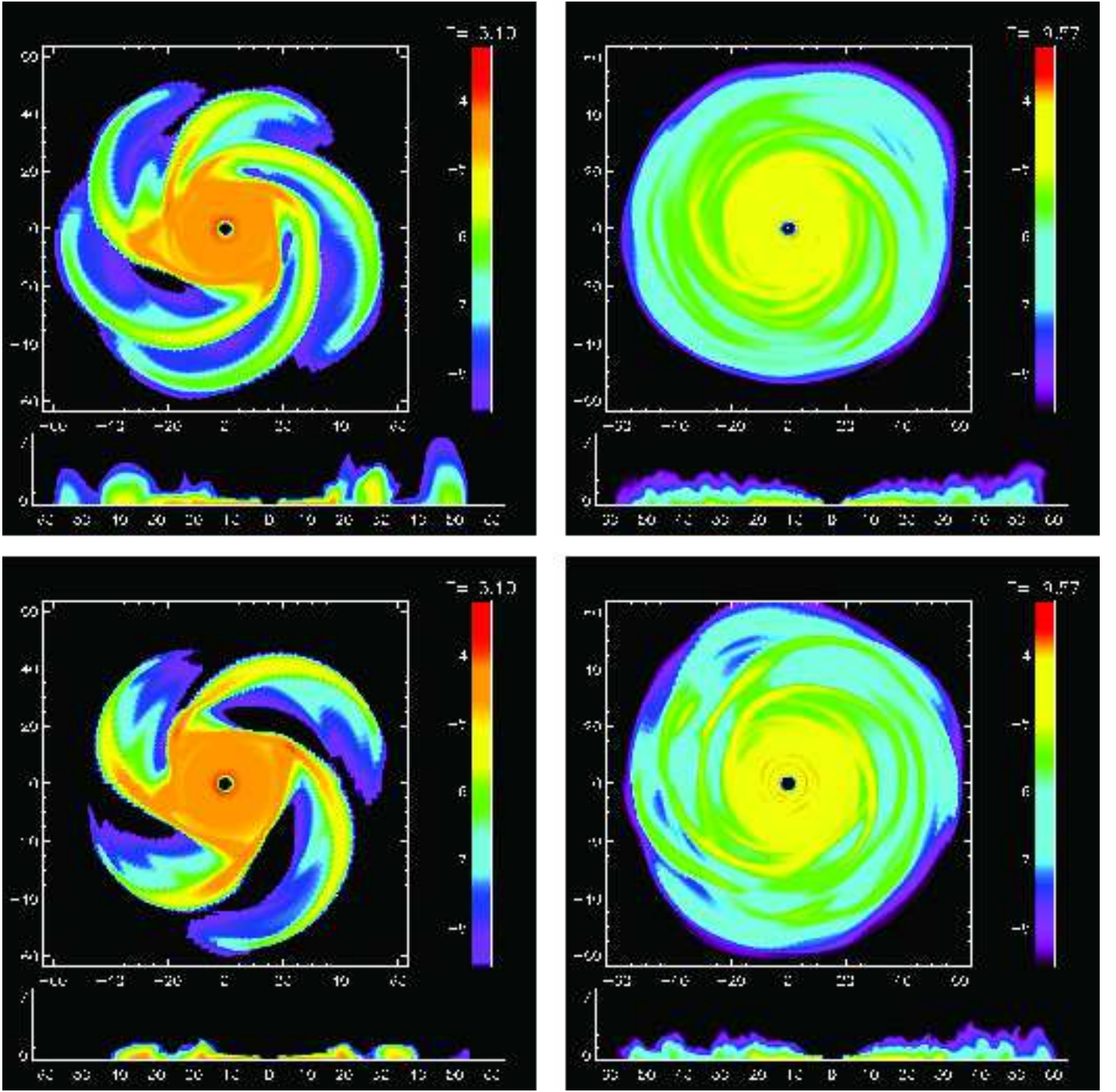}
\caption{Midplane densities in logarithmic scale of the indirect (top) and fixed (bottom) simulations. The axes have units of AU and the time in the upper right is given in ORPs. The small white x in the indirect panels is the position of the system center of mass (disk plus star). The left panels show the saturation of the initial growth in the burst phase. The right panels are representative of the asymptotic behavior.}
\label{SMfinalcomp}
\end{figure*}

\section{Methodology}
\label{sec:method}

\subsection{The IUHG hydrodynamics code}
\label{sec:numerics}

The two disk simulations in this paper are evolved using the standard IUHG 3D hydrodynamics code \citep{pickett2003,mejia2005}. This code solves the equations of hydrodynamics in conservative form on an evenly spaced Eulerian cylindrical grid ($\varpi$, $\phi$, $z$) to second order in space and time. For these calculations, the equation of state is that of an ideal gas with a ratio of specific heats $\gamma = 5/ 3$. The rotation axis is in the $z$-direction, reflection symmetry is assumed about the equatorial plane, and the cell widths in the $\varpi$ and $z$-directions are the same. Shocks are mediated through the inclusion of von Neumann-Winkler artificial bulk viscosity terms in the momentum and energy equations \citep{pickettphd1995}, and free outflow boundary conditions are assumed along the top, outer, and inner (central hole) edges of the grid. The ($\varpi$,$\phi$,$z$)-directions will hereafter be referred to as the cylindrically radial, azimuthal, and vertical directions, respectively. When needed, the spherical radius from grid center is represented by $r$. Cooling is introduced into the energy equation as a local volumetric cooling rate $\Lambda = \epsilon/t_{\mathrm{cool}}$, where $\epsilon$ is the internal energy density and $t_{\mathrm{cool}}$ is a constant time scale on which the disk is assumed to be losing thermal energy due to radiation. 

The potential whose gradient yields the gravitational source terms in the momentum equation of the disk is made up of several components, namely, the disk component, the stellar component, and the indirect component. So 
\begin{equation}
\Phi_{\mathrm{tot}}=\Phi_{\mathrm{disk}} + \Phi_{\mathrm{star}}+\Phi_{\mathrm{ind}}. \label{poteq}
\end{equation} 
$\Phi_{\mathrm{disk}}$ is calculated by first determining the disk boundary potential by multipole expansion of spherical harmonics up to $l=|m|=10$. Once the boundary conditions have been obtained, the density data is Fourier transformed in the $\phi$-direction. Each of the Fourier components yields a 2D boundary value problem which can be cast into a block-tridiagonal matrix and  solved using cyclic reduction \citep{tohline1980}. The solution in Fourier space is then transformed back into real space. $\Phi_{\mathrm{star}}$ is $-GM_{\ast}/r$, and $\Phi_{\mathrm{ind}}$ is described in the next section. For the comparison calculation in \citet{mejia2005}, $\Phi_{\mathrm{ind}}$ is zero. Although the star remains at the grid center in both simulations that we discuss, we refer to the \citet{mejia2005} simulation without the indirect potential as the ``fixed star'' case and the simulation with the indirect potential as the ``indirect case''.  

In our earlier constant $t_{\mathrm{cool}}$ simulations with 256 or 512 radial grid cells that have a fixed central stellar potential, the disk's COM moved at most only a few radial cells away from the grid center over many disk rotations due to numerical errors.  Nevertheless, even if fixing the star at the center of the inertial frame does not lead to pathological behavior, it also does not capture the full interaction between the star and disk, including possible feedback between stellar motion and growth of spiral modes \citep[e.g.,][]{adams1989,shu1990}. 

\subsection{Indirect potential}
\label{sec:indirect}

To model the stellar motion explicitly, we use the indirect potential method \citep[e.g.,][]{nelson2000b}. The grid is now considered to be the accelerated reference frame of the star, with the practical benefit that the star remains fixed at the grid center. The acceleration of the star-centered reference frame due to gravitational forces can be included into the fluid equation of motion as an additional gradient of a potential, termed the {\sl indirect potential}. As in \citeauthor{nelson2000b}, the indirect potential at some point in the disk is given by
\begin{equation}
\Phi_{\mathrm{ind}} = G\int_{V^\prime} \frac{\mathrm{d} m(\boldsymbol{r^\prime})}{r^{\prime3}}\boldsymbol{r}\cdot\boldsymbol{r^\prime},\label{phiind}
\end{equation}
where the integration over the primed coordinates extends over the whole computational grid. In practice, we compute
\begin{equation}
\int_{V^\prime}\frac{\mathrm{d} m(\boldsymbol{r^\prime})}{r^{\prime3}} \boldsymbol{r^\prime}
\end{equation} 
and then 
\begin{equation}
G\boldsymbol{r} \cdot \int_{V^\prime}\frac{\mathrm{d} m(\boldsymbol{r^\prime})}{r^{\prime3}} \boldsymbol{r^\prime}
\end{equation}
for each cell. Of course, this turns out to have much less computational burden than computing the entire integral in equation \eqref{phiind} for each cell.

\subsection{Initial model}
\label{sec:initial}
For this comparison, we use the same initial disk as \cite{mejia2005}, except that we rescale the physical parameters so that $M_{\mathrm{star}} = 1$ $\mathrm{M_\odot}$ and $M_{\mathrm{disk}} = 0.14$ $\mathrm{M_\odot}$. The \citeauthor{mejia2005} simulation results are also appropriately rescaled in all comparisons presented here. The disk extends from 2.3 to 40 AU with a surface density $\Sigma \propto \varpi^{-1/2}$. The computational gird has a central hole with a radius of 1.6 AU in which the star resides, and the 40 AU initial disk outer radius is at radial zone 242. The new run with the indirect potential is started from the same axisymmetric disk after applying the same $0.01\%$ amplitude random cell-to-cell density perturbation used by \citeauthor{mejia2005}. As in \citeauthor{mejia2005}, the constant $t_{\mathrm{cool}}$ is set equal to 2 Outer Rotation Periods or ORPs. An ORP is defined as the initial rotation period, about 180 yr, at radial zone 200 ($\approx~$33 AU). The cylindrical grid in ($\varpi$,$\phi$,$z$) is initially 256x128x32. An additional 256 radial zones are added at about 5 ORPs to extend the grid to about $\varpi =$ 85 AU once GIs cause the disk to expand radially off the initial computational grid. The initial $Q$-distribution has a minimum of about 1.5 at $\varpi \approx$ 34 AU, and so the disk is only marginally stable at time $t = 0$. The indirect potential simulation is run for a total of about 19.6 ORPS or 3,500 yr. 

In the \citet{mejia2005} simulation, the fixed initial central ``star'' is a radially extended oblate mass distribution inside the central disk hole. When fitting the Mej\'ia star by a point mass for the indirect case, an error of 0.1\% is made in the star's mass. An additional error in the central force due to the nonsphericity of the Mej\'ia star is about 0.1\% at 11 AU and falls off quickly outside that radius. These discrepancies are two orders of magnitude less than the tens of percent differences reported for the GI behavior outside 10 AU and should be unimportant.

\section{Results}
\label{sec:results}

\subsection{Overall evolution}\label{overall}

\subsubsection{Evolutionary phases}\label{results:phases}

Qualitatively, the overall outcomes are fairly similar. The disk simulated with the indirect potential goes through the same four phases as the disk in the fixed star case \citep{mejia2005}, namely {\sl axisymmetric cooling}, the onset of GIs in a {\sl burst}, and a {\sl transition} to a final quasi-steady {\sl asymptotic} phase.  The dividing times between these phases are also roughly the same. The onset of the burst phase occurs around 4.5 ORPs with the initial burst being predominately in discrete four to six-armed modes. The transition phase begins near 7 ORPs and continues until the start of the quasi-steady asymptotic phase around 11 to 12 ORPs, after which heating by GI activity balances cooling on average throughout the GI-active region and instability of a large number of interacting spiral waves is sustained in an approximate steady state \citep[see also][]{gammie2001}.

\subsubsection{Final state}\label{results:final}

\begin{figure}
\includegraphics[width=3.5in]{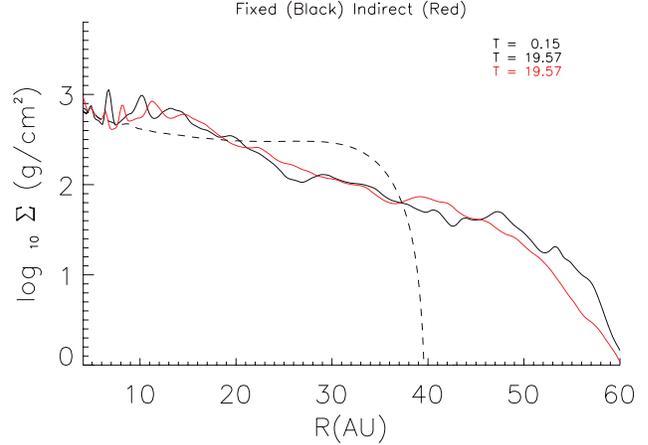}
\caption{Surface densities in logarithmic scale of the indirect (red) and fixed (black) simulations.}
\label{SMsurfcomp}
\end{figure}

The right-hand panels in Figure \ref{SMfinalcomp} offer a comparison of the final midplane densities for the disks. The structures in the indirect potential simulation appear to be a bit less sharply defined, probably owing to somewhat weaker GI amplitudes, as discussed in \S\ref{results:asymptotic}. One can also see that the system COM, indicated by the white x in Figure \ref{SMfinalcomp}, has remained within the central hole of the grid. Although this motion is not large from the disk's perspective, it may be of observational interest (see \S\ref{discussion:star}). Additionally, as can be seen in Figure \ref{SMsurfcomp}, the final azimuthally-averaged surface density profiles show very little difference, except that the indirect potential disk is slightly less extended radially. Assuming the surface density profile is a power law $\Sigma \propto r^{-q}$, a least squares fit for $q$ between 15 and 40 AU gives 2.33 and 2.31 for the indirect potential and fixed star cases, respectively. Also, the Toomre $Q$-values are similar in the asymptotic phase and average to about 1.2 to 1.3 over the 15 to 40 AU region. 

\subsubsection{Differences in the burst phase}\label{results:burst}

Despite the overall similarities, the GI structure for the indirect case exhibits some interesting differences. These can best be illustrated quantitatively by examining the overall amplitudes $A_m$ of $\sin(m\phi)$ and $\cos(m\phi)$ terms in a Fourier decomposition of the density in the azimuthal direction. We compute the Fourier components as in \citet{imamura2000} such that 
\begin{equation}
A_{m} =\frac{(a^2_{m}+b^2_{m})^{1/2}}{\pi\int\rho_0 r dr dz},
\end{equation}
where
\begin{equation}
a_m = \int \rho \cos (m\phi) r dr dz d\phi
\end{equation}
and
\begin{equation}
b_m = \int \rho \sin (m\phi) r dr dz d\phi .
\end{equation}
Here $\rho_0$ is the axisymmetric component of the density, and the integrals extend over the computational grid. We typically sample the value of the components 100 times per ORP and, for some purposes, average them over a time interval, which is typically several ORPs in the asymptotic phase. 

As shown in Figure \ref{SMfinalcomp}, there is a qualitative difference in dominant mode during the burst. For both cases, analyses of $A_m(t)$ plots reveal exponential growth in the linear regime at similar rates for $m = 3$ to 6 spiral waves centered near 30 AU, the region where $Q$ is initially minimum. On the other hand, the $m = 4$ and 5 waves play very different roles in the two cases. $A_4$ grows first and is dominant for the fixed case; growth in $A_5$ is substantially delayed. On the other hand, $A_5$ and $A_6$ dominate growth for the indirect case, with $m = 4$ remaining unimportant until amplitudes become nonlinear. All spiral waves grow from random uncorrelated noise imposed initially in both cases that displaces the COM from the grid center. In the fixed case, where deviations of the star from the COM are not handled physically, an ordering in the small ($\sim 10^{-4}$) amplitudes of $m = 4, 5,$ and 6 develops within a small fraction of an ORP such that $m = 4$ has a head start in its linear growth phase and reaches nonlinear amplitude about a third of an ORP before the other modes. On the other hand, in the indirect case, where the COM is treated properly, the early values of $A_2$ to $A_6$ are more nearly the same, as they should be for white noise. Then $m = 5$, followed closely by $m = 6$, dominates the linear growth. During the linear growth phase for both cases, $A_1 \approx A_3A_4 + A_4A_5 + A_5A_6$, as expected if $m = 1$ results from the nonlinear interaction of the $m = 3$ to 6 modes and is not itself an independent unstable mode \citep[e.g.,][]{laughlin1996}. As discussed in \S\ref{SMdetailed}, the motion of the star with respect to the COM during the burst of the indirect case appears to be a response to this nonlinear interaction and should be sensitive in detail to the modes involved. Because our initial conditions are arbitrary and the subsequent evolution is not substantively different, the qualitative difference in dominant mode during linear growth does not undermine the general conclusions about disk evolution drawn from simulations with fixed stars in earlier IUHG papers. On the other hand, the significant displacement of the star during the burst and subsequent phases is a potentially interesting and observable consequence (see \S\ref{discussion:star}).

\subsubsection{Differences in the asymptotic phase}\label{results:asymptotic}

As shown in Figure \ref{SMamspec}, all the $A_m$ values are lower (about 0.05 to 0.2 dex in $A_m$, or on the order of tens of percent) for the indirect case during the asymptotic phase. The bigger differences appear to be in the higher-order Fourier components. Interestingly, although the indirect case allows for the possibility of  SLING amplification, $A_1$ is in fact measurably {\sl lower} in the indirect case. The ``error bars" on each of the points indicate the rms fluctuations over the sampling interval. It should be noted that the number of epochs used to compute the average $A_m$ and their fluctuations is different for each of the runs, namely 801 and 504 for the indirect and fixed star cases, respectively. This is due to sampling data for a fixed number of time steps, even though the time step length actually differed between the calcualtions. Because the number of epochs sampled is large in both cases, we do not think this causes significant differences in the average $A_m$ or rms fluctuations. 

As a measure of the total amplitude of non-axisymmetric density structure, we can sum $A_m$ over $m$ from 1 to 64. For the time period between 4.5 and 11 ORPs, representing the burst and transition phases, the peak summed amplitude is 5.8 in the indirect case as opposed to 7.6 in the fixed star case. The average of the summed amplitudes over this time interval is also lower, at 2.1 for the indirect case compared to 2.7 in the fixed star case. This trend continues throughout the asymptotic phase with the mean value of the summed amplitudes being 2.2 in the indirect case compared to 2.7 in the fixed case averaged over the interval from 12 to 19.5 ORPs. Figure \ref{SMamspec} shows the full range of $m$ values averaged over this time.

\subsubsection{Mass transport}\label{results:transport}

Figure \ref{SMmdots} compares the disk mass transport rates during both the burst phase and the asymptotic phase, as determined by changes in the distribution of disk mass with respect to cylindrical radius $\varpi$ measured from the star. Overall, we see that the addition of stellar motion does not affect the general outcome of GI activity for mass transport, even though it does produce a measurable weakening of the GI amplitudes. In both cases, typical radial inflow rates in the disks during the burst and transition phase are $\sim  3$ to $7\times10^{-6} M_{\odot} {\rm yr}^{-1}$ over 15 to 35 AU. The inflow is harder to characterize for the asymptotic phase but crudely averages to something like $\sim 7\times10^{-7} M_{\odot} {\rm yr}^{-1}$ over 10 to 35 or 40 AU in both cases. \citet{mejia2005} have already demonstrated how much fluctuation there is in mass inflow due to GIs (see the light grey curves in their Figure 3). 

\subsection{Stellar motion}\label{SMdetailed}

With the indirect potential, we can infer the physical motion of the star by determining the position of the system COM on our grid as a function of time. Figure \ref{SMstarmotion} shows the COM's equatorial plane trajectory in the star's reference frame and the distance $\varpi_{\rm com}(t)$ between the star and COM as a function of time for the duration of the simulation. In Figure \ref{SMstarmotion}, the black depicts the COM motion from 0 to 12 ORPS, i.e., through the burst and transition phases, while the red shows the asymptotic phase for $t >$12 ORPs. The star does not become significantly displaced from the system COM until the burst commences at 4.5 AU and achieves its largest excursion from the origin, $\sim$ 0.25 AU, near the end of the burst at about 7 ORPs. 

As discussed in Section \ref{results:burst}, there does not appear to be sustained growth of an independent $m = 1$ mode through a feedback loop. The initial displacement of the star during the burst appears to be due to the nonlinear interaction of odd and even spiral waves with $m = 3$ to 6. As can be seen in Figure \ref{SMamtrcom} there is a clear correlation between the exponential growth of the $A_m$ and the displacement of the COM. The COM begins with a very small displacement due to the random perturbation in the initial disk. The COM remains at this small radius until about a half ORP after the onset of exponential growth in the $A_m$s. The time lag is probably due to the need for a wave to propagate to the edge of the disk before the COM displaces (see step 3 in Section 1c of \citet{shu1990}). Once the disk enters the asymptotic phase, the COM excursion decreases in amplitude but remains non-negligible. From 12 to 19.5 ORPs the mean distance from the star is 0.09 AU, while the average distance of the COM from the star is about 0.11 AU over the entire time period between 4 and 19.5 ORPs.

\begin{figure}
\includegraphics[width=3.5in]{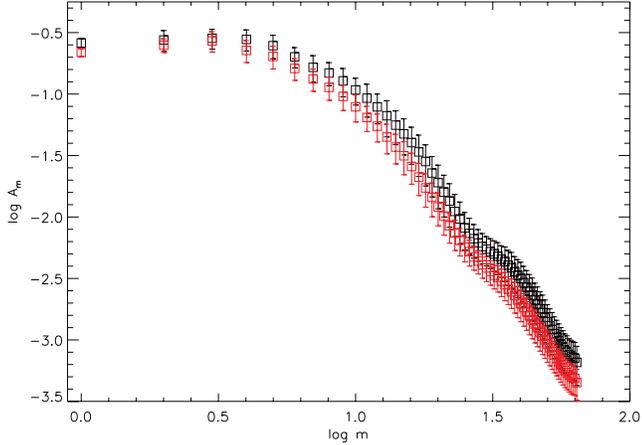}
\caption{Comparison of indirect (red) and fixed (black) $A_m$ values for each $m$ averaged from 12 to 19.5 ORPs. The error bars represent the RMS of the fluctuations over this time period.}
\label{SMamspec}
\end{figure}

\begin{figure}
\includegraphics[width=3.5in]{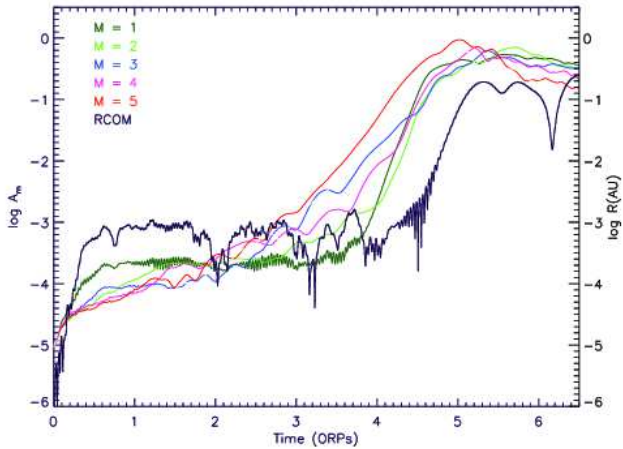}
\caption{$A_m(t)$ and $\varpi_{\rm com}(t)$ for the initial phase of the indirect simulation. The vertical $A_m$ scale is on the left; the COM scale is on the right.}
\label{SMamtrcom}
\end{figure}

\begin{figure}
\includegraphics[width=3.5in]{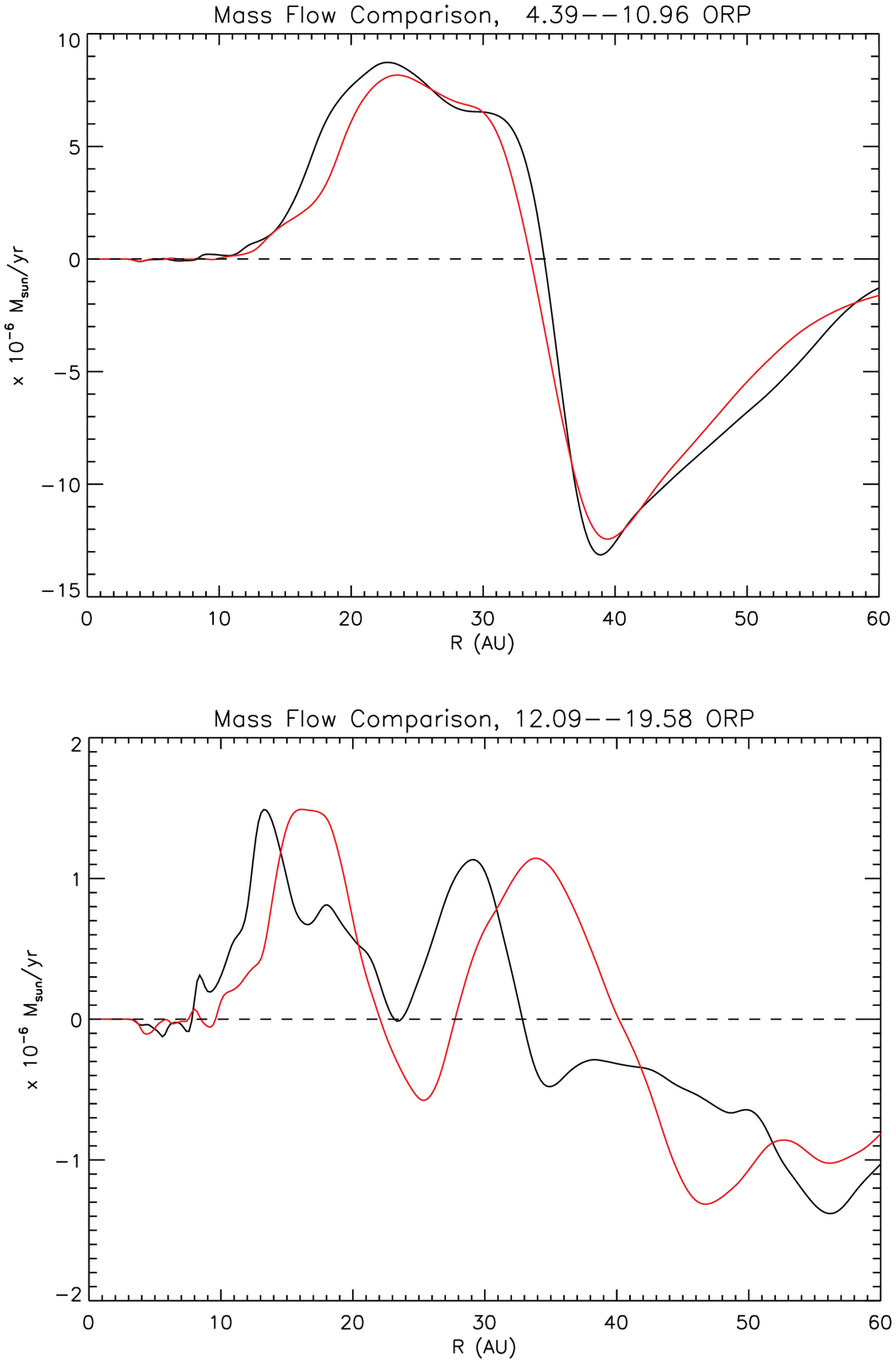}
\caption{Mass transport rates in solar masses per year plotted during the burst (top) and asymtotic (bottom) phases. The horizontal axis is given in AU and the red curve is the indirect simulation, the black curve is the fixed simulation and the dashed line represents no mass flow.}
\label{SMmdots}
\end{figure}

To determine the periodicity of the COM motion, and hence, by reflex, of the star, we construct Lomb-Scargle periodograms \citep{lomb1975,scargle1982} for the equatorial plane COM motion in $\varpi_{\rm com}$ and in Cartesian $x$ and $y$ coordinates. Figure \ref{SMcomperiod} shows the power spectrum for $x$ measured between 12 and 19.5 ORPs; the $y$ periodogram (not shown) is similar. If we assume that the most powerful signal is due to the star's primary orbit motion, we get a period of 0.80 ORP (about 140 yr) for both $x$ and $y$. The periodogram for $\varpi_{\rm com}$ gives consistent results but suggests a slightly shorter period of 0.72 ORP (about 130 yr). One expects the $x$ and $y$ periods to agree, while the $\varpi_{\rm com}$ period should be shifted somewhat due to apsidal precession, as observed. Several statistical tests were done to verify the significance of the signals, including changes in the sampling rate of the COM position (see Michael 2010, Ph.D. dissertation). The signal is robust, and one also observes a similar period just by counting peaks in the $\varpi_{\rm com}(t)$ plot over the same time interval. As in \citet{rice2003a}, there are significant higher-order motions superposed on the basic COM wobble.

During the burst phase, the period of the COM motion appears to be somewhat longer, closer to 1 ORP, in rough agreement with the pattern period of the predominant five-armed mode (see the top left panel of Figure \ref{SMfinalcomp}). Although the burst is dominated by the $m = 5$ pattern, both odd and even modes with similar pattern periods grow together. The resulting $m = 1$ distortion of the disk due to the nonlinear interaction of these modes appears to kick the star into orbit around the system COM with a period similar to the pattern period of the growing spiral waves. This is consistent with the following simple estimate. The star's largest excursion from the COM is 0.25 AU, while corotation for the bursting spiral waves is at about 30 AU. So the star's motion implies a net mass asymmetry in the disk of about $0.25 M_{\mathrm star}/30 \approx 0.008 M_{\odot}$ (about eight Jupiter masses) at about 30 AU on the other side of the COM. The mass in the disk between the Lindblad resonances of the dominant $m = 5$ mode during the burst is approximately $0.03 M_{\odot}$. So just a fraction of the mass (a bit more than one spiral arm's worth) involved in the interacting odd ($m = 3$ and 5) and even ($m = 4$ and 6) waves is sufficient to displace the star by the amount observed.

\begin{figure}
\includegraphics[width=3.5in]{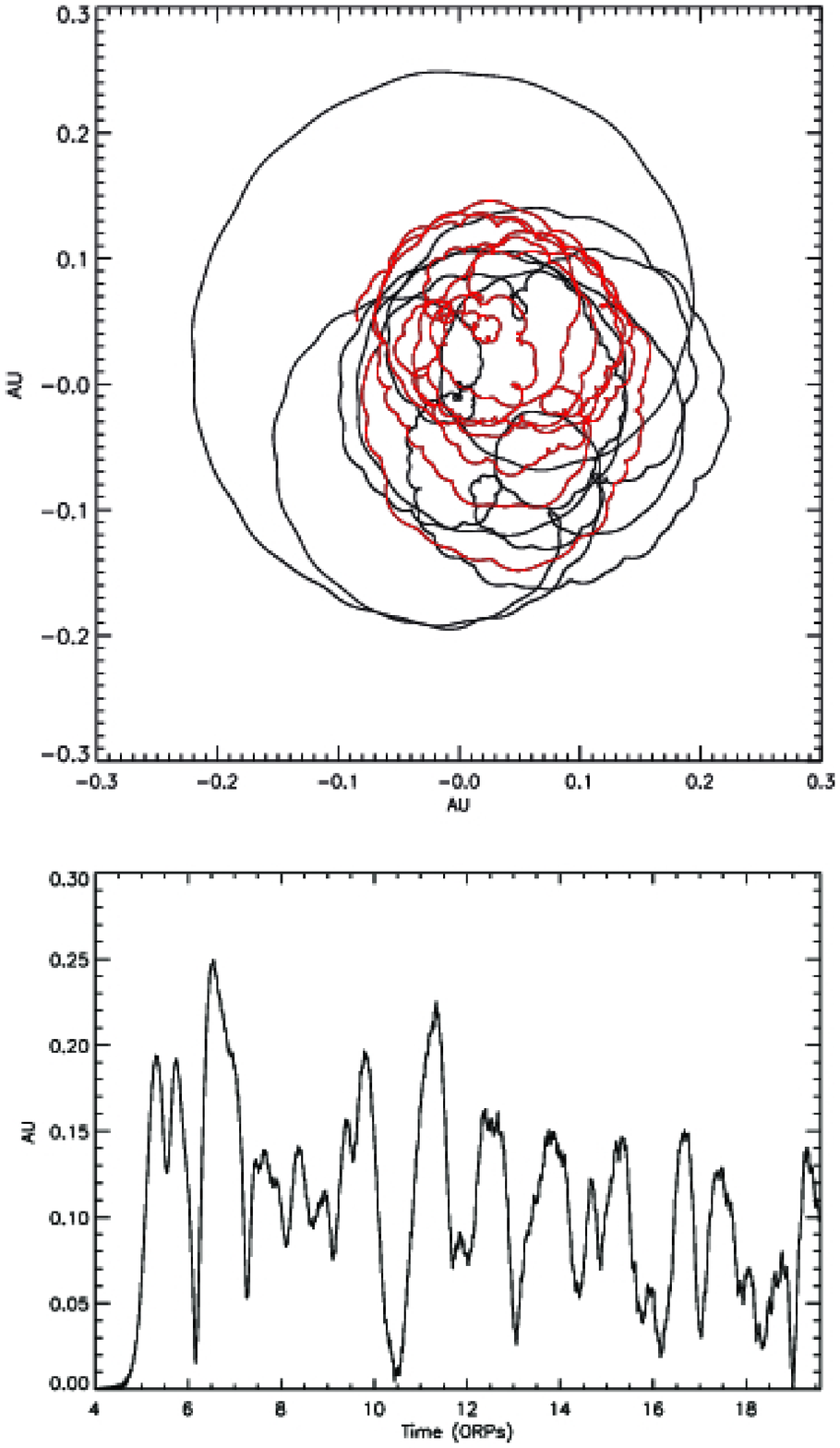}
\caption{System COM motion. The top plot shows the COM trajectory in the equatorial plane, and both axes are labelled in AU. Black represents motion prior to 12 ORPs, and red after. The bottom plot shows the system COM radial excursion for the duration of the simulation, radial excursion labelled in AU and time in ORPs.}
\label{SMstarmotion}
\end{figure}

\begin{figure}
\includegraphics[width=3.5in]{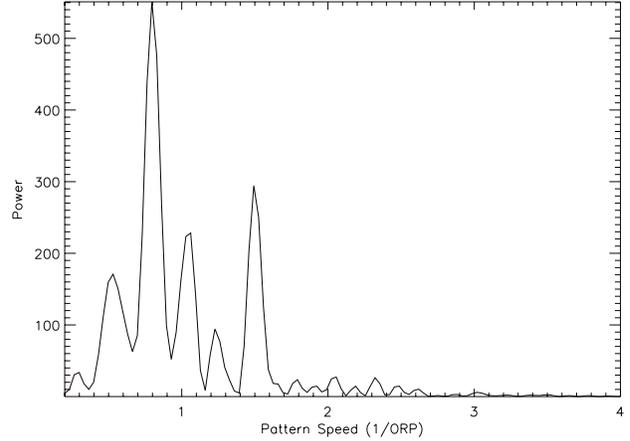}
\caption{Lomb-Scargle periodogram using data from 12 to 19.5 ORPs for the $x$ motion of the system COM. The power on the vertical axis is in arbitrary units.}
\label{SMcomperiod}
\end{figure}

\begin{figure}
\centering
\unitlength1in
\includegraphics[width=3.5in]{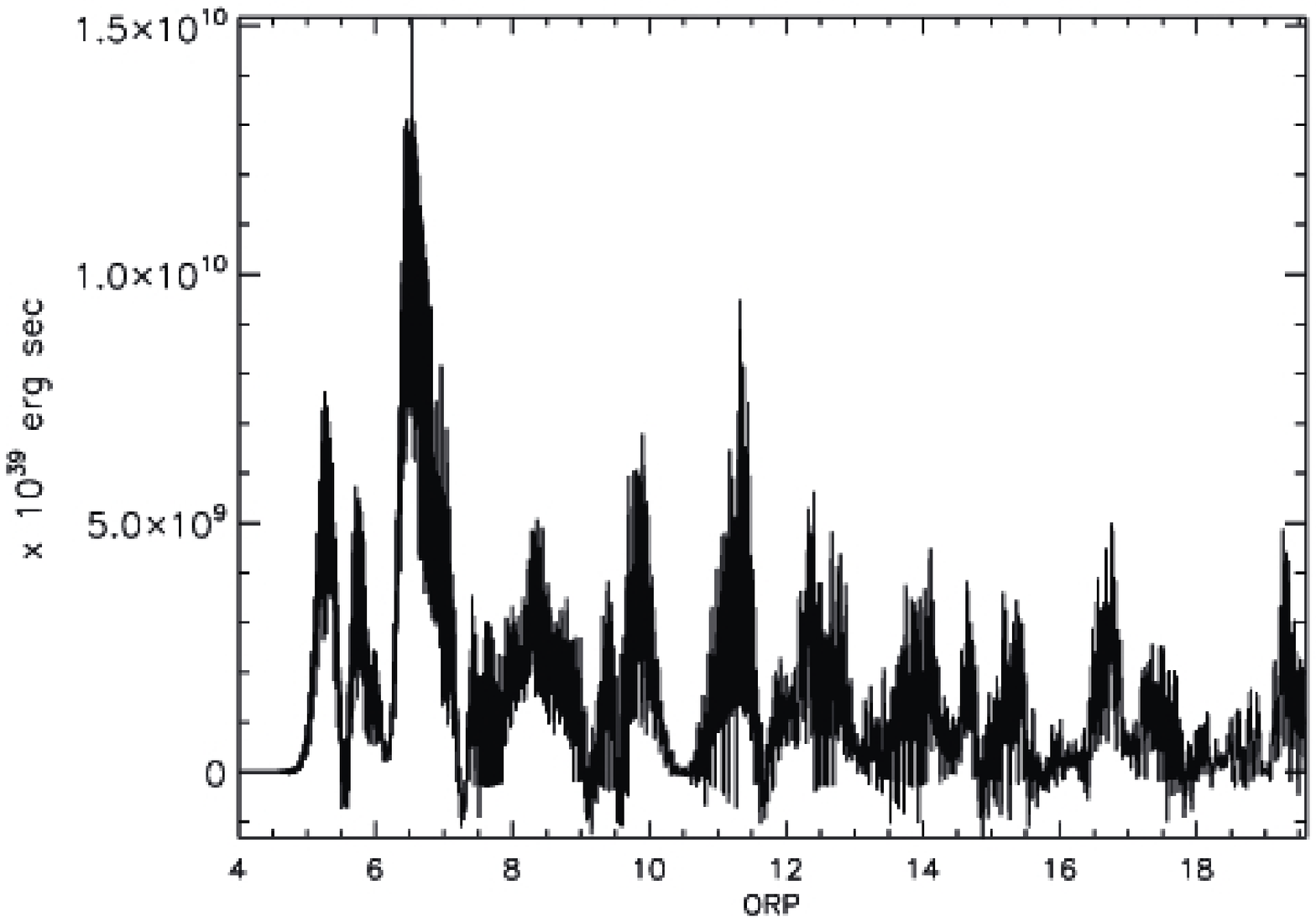}
\caption{Plot of the stellar angular momentum as a function of time as determined from the motion of the system COM.}
\label{SMcomtorque}
\end{figure}

The significant motions of the star in the inertial frame show that an appreciable amount of angular momentum is transfered between the disk and the star. Figure \ref{SMcomtorque} plots the star's angular momentum produced by the star/disk interactions, as calculated by taking a five point numerical time derivative of the system COM position vector to get a velocity vector. The COM position is determined by a numerical integral over the grid, and the velocities are then determined by numerical derivative, so numerical errors are amplified and the result is fairly noisy. We have tried to estimate gravitational torques by doing a numerical derivative of the curve in Figure \ref{SMcomtorque}, but the results have even more noise, and we are not sure how much of the fluctuation is numerical or due to the gravitoturbulence in the disk. The average torque during the asymptotic phase is a few $\times10^{38}$ergs, with a suggestion that instantaneous torques can be much larger or smaller. 

\subsection{Analysis of the disk structure} \label{SMdisk}

In Figure \ref{SMamspec}, we compared the GI amplitudes in various Fourier components for the two simulations during the asymptotic phase. Another important characteristic of $m$-armed spiral waves is the coherence of their patterns. As explained in \citet{mejia2005}, we can search for patterns that are coherent over a range of radii by creating power spectra at all radii for ${\rm cos}(\phi_m)$, where $\phi_m$ is the phase of the $m$-th Fourier component of the azimuthal density structure. We restrict the analysis to the equatorial plane. Figure \ref{SMm1period} compares periodograms for $m = 1$, 2, and 3 for the two simulations during the asymptotic phase. Within each diagram, the relative power, as denoted by the color scale, is a measure of the coherence of the patterns detected at each radius. The apparent difference in color scale between plots for the two simulations is an artifact of somewhat different sampling rates with time and has no significance. What matters are the strong vertical strips, which indicate periods that are present over a range of radii, i.e., spatially coherent wave patterns. Curves representing the approximate locations of the outer Lindblad resonance (OLR), corotation (CR), and the inner Lindblad resonance (ILR) for a given pattern period $P_{\rm pat}$ are shown. The curves are a bit ragged due to the use of azimuthal averages and/or numerical derivatives to compute these radii numerically. As pointed out in \citet{mejia2005}, coherent patterns tend to be confined between their Lindblad resonances.

\subsubsection{One-armed structure}\label{SMm1}

\begin{figure*}
\includegraphics[height=9in]{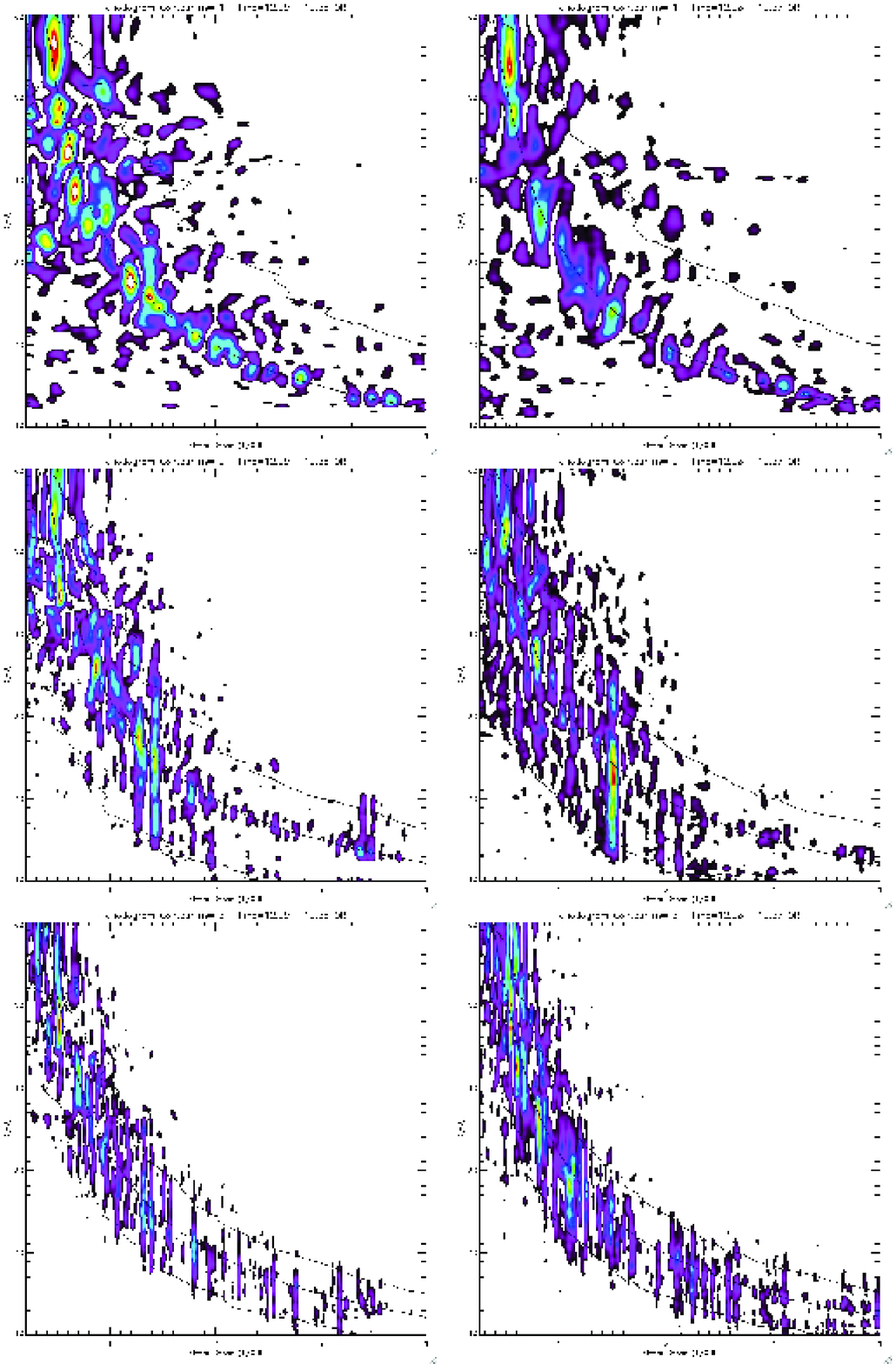}
\caption{The $m=1,2,$ and 3 phase periodograms measured using data from 12 to 19.5 ORPs. The color scale represents relative power with purple being the smallest  and red/white being the largest. In this plot, relative power is the important measure; the absolute numbers are in arbitrary units. The fixed star case is on the left, and the indirect case on the right.}
\label{SMm1period}
\end{figure*}

In Figure \ref{SMm1period}, $m = 1$ does not exhibit any stronger radially coherent patterns in the indirect simulation. In fact, except perhaps for some slow-moving distortion of the outer disk between 50 and 60 AU, where there is little mass, neither simulation shows any vertical stripes of power in $m = 1$. There are no coherent global one-armed spiral pattens in the disks. Because SLING is most likely to amplify one-armed modes \citep{adams1989,shu1990}, this suggests that sustained SLING amplification is not present during the asymptotic phase of either simulation. Using the methods described in \citet{boley2006} and \citet{cai2008}, Figure \ref{SMdisktorque} compares the gravitational torque of the outer disk on the inner disk for several $m$-values at all $\varpi$ during the asymptotic phase. For both simulations, these disk internal gravitational stresses are measured using cylindrical radii centered on the star. Figure \ref{SMdisktorque} shows negligible contribution to the torque from the $m = 1$ component in both simulations.

\subsubsection{Disk/star interaction} \label{SMdiskstar}
\begin{table}
\centering
\caption[Pattern Speeds of Periodogram Features]{Pattern Speeds of Prominent Periodogram Features}
\begin{tabular}{ccccc}
\hline
Signal Measured&\multicolumn{4}{c}{Pattern Speeds (1/ORP)}\\
\hline\hline
COM $x$-motion&0.55&0.80&1.05&1.50\\
COM $y$-motion&0.51&0.80&1.02&1.50\\
$m = 1$ &0.65&0.85&---&1.50\\
$m = 2$ &0.50&0.80&---&1.50\\
$m = 3$ &0.59&0.80&1.05&1.50\\
\hline
\label{SMperiodtable}
\end{tabular}
\end{table}

\begin{figure}
\centering
\includegraphics[width=3.5in]{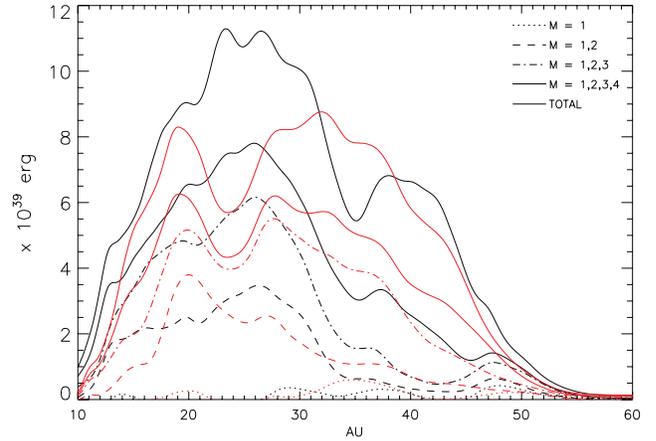}
\caption{Torque profiles averaged from 12 to 19.5 ORPs for the indirect (red) and fixed (black) simulations.}
\label{SMdisktorque}
\end{figure}

Although the motion of the star does not induce sustained growth of coherent $m = 1$ structure, the star clearly has a strong interaction with the disk. Table \ref{SMperiodtable} compares the strongest periods in the COM $x$ and $y$ periodograms and the strongest stripes of power in the disk periodograms. The periodicity in the $x$ and $y$ COM motion corresponds to maxima in $m = 1, 2, \mathrm{and }~3$ periodograms at pattern speeds near $0.5, 0.8, \mathrm{and }~1.50$/ORP. The most striking features in the $m = 2~\mathrm{ and }~3$ periodograms are those at pattern speeds of 1.50/ORP and 1.05/ORP, respectively. Figure \ref{SMdisktorque} shows the effect of this star/disk interaction on the disk torque.  Clearly, the disk torque is decreased in the indirect simulation when compared to the fixed simulation, especially near 24 AU, which corresponds to the corotation radius of a pattern with a pattern speed of about 1.50/ORP. Although the interpretation is unclear, we also note that the torque maximum at 19 AU roughly corresponds to the inner Linblad resonance of a coherent $m = 3$ pattern with a pattern speed of near 1.5/ORP. Moreover, the decrease in disk torque centered near 24 AU is comparable to the torque on the star. The time averaged torque from 12 to 19.5 ORPs on the star is about $5 \times10^{38}$ ergs, while the difference in disk torques averaged over the same time interval and averaged from 10 to 60 AU is about $9\times10^{38}$ ergs. This is essentially agreement within the accuracy to which these averaged torques can be measured. Apparently, this amount of torque, which is confined to the disk in the fixed star case, is effectively transferred to the star when the star is allowed to move.

\begin{figure}
\includegraphics[width=3.5in]{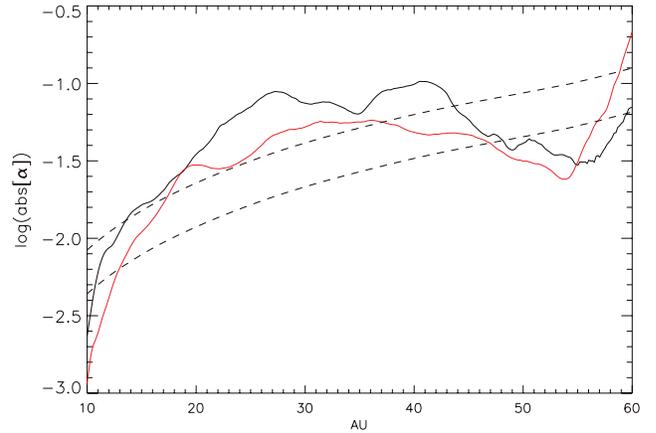}
\caption{Effective Shakura-Sunyaev $\alpha$-values computed for the fixed star (black curve) and indirect (red curve) simulations averaged over the asymptotic phase from 12.5 to 19 ORPS. Shown for comparison are curves predicted by equation (13) in \citet{gammie2001} with $t_{cool} = 2$ ORP. The upper curve assumes a vetical structure dominated by disk self-gravity; the lower curve assumed the vertical gravity is due to the star alone. For $Q \approx 1.3$, we expect the disk to be significatly self-gravitating.}
\label{SMalphas}
\end{figure}

Even though these differences in the disk structure introduced by the star's motion are measurable in the torque profiles and periodograms, the overall mass transport in the disk is largely unaffected as can be seen in the mass transport rates of Figure \ref{SMmdots}. Another metric of mass transport is the effective gravitational $\alpha$, which is plotted for both the fixed and indirect simulations in Figure \ref{SMalphas}. Figure \ref{SMalphas} also shows the predicted \citet{gammie2001} values for fully self-gravitating and non-self-gravitating disks for $t_{\mathrm{cool}} = 2$ ORP when the disks are in a local balance between radiative cooling and heating by GI activity. Although the indirect potential $\alpha$ is somewhat smaller than the fixed-star $\alpha$, it follows the overall trend of the fixed-star $\alpha$. The decrease in $\alpha$ is consistent with the decrease in torque, becasue these effective $\alpha$-values are computed from the torques \citep[see][]{boley2006}. Notice that the inclusion of stellar motion brings the measured $\alpha$ somewhat closer to values predicted by the local balance assumption \citep{gammie2001}, which are also shown on the plot. The agreement between the $\alpha$-values from indirect simulation and the local treatment are closest in the region around the corotation of the strongest $m = 2$ and 3 patterns in Figure \ref{SMm1period}. It was argued by \citet{balbus1999} that agreement between global and local energetic behavior of GIs was most likely to occur near corotation.

\section{Discussion}
\label{sec:compare}

\subsection{The effect of star/disk interactions}\label{discussion:GIs}

We have found that the inclusion of stellar motion in our 3D radiative hydrodynamic simulations has a measurable, but not drastic effect on the evolution of a gravitationally unstable protoplanetary disk. Gross features such as surface density profile, $A_m$ values, and mass transport rates remain unchanged within tens of percent. On the other hand, the motion of the central star is not negligible, and there is a relationship between changes in the disk self-torque and the star/disk interaction. 

During the burst phase, the freedom of the star to move qualitatively changes the spiral modes that dominate at the initiation of GI activity, and the stellar motion reflects the orbit period at the CR of the growing modes. During the asymptotic phase, there is also a strong correlation between the quasi-periodic motions of the star and periodicities evident in the $m = 1-3$ structure, as seen by comparing Figures \ref{SMcomperiod} and \ref{SMm1period} with Table \ref{SMperiodtable}. Despite the complex interaction between the disk and star that gives rise to this correlation, measures of GI-activity in the disk during the asymptotic phase only tend to be reduced by at most about tens of percent. Overall, it appears that the net effect of the star/disk interaction is to introduce a new degree of freedom to the system which allows some of the torque that previously went into disk self-torque to be applied instead to the star/disk interaction. This, in turn, slightly lowers the $A_m$ values, disk self-torques, and resultant $\alpha$. The star/disk torque interaction transfers a significant amount of angular momentum into orbital motion of the star. Further study is required to determine the sensitivity of this interaction to various simulation conditions, such as cooling time (and thereby GI strength), initial surface density profile, and disk-to-star mass ratio.

Figure \ref{SMdisktorque}, which divides the contribution to the disk torque into its various Fourier components during the asymptotic phase, shows negligible contribution from $m = 1$ or one-armed disturbances, which would be expected to dominate if SLING were effective in our disk. Analysis of the $A_m(t)$ plots during the burst (not illustrated here) also shows that $m = 1$ only grows as a result of the nonlinear interaction of spiral modes with higher $m$ as the GI activity begins.  There are two possible reasons for failure to detect evidence for a one-armed instability: 1) Although a dynamic process, the growth times for one-armed instabilities, whether driven by SLING or not, tend to be several to many outer rotation periods \citep{adams1989,shu1990}, because one-armed instabilities involve a reflection from the outer boundary of the disk. Growth times become especially long for disk-to-star mass ratios of order 0.1, as illustrated by the low growth rates in Figure 4a of \citet{noh1992}. Our simulations probably do not last long enough for one-armed instabilities to grow. 2) SLING amplification in particular probably does not occur because our disk-to-total-mass ratio is too low \citep{shu1990,noh1992}. 

\subsection{Observable consequences of stellar motion}\label{discussion:star}

The stellar displacement from the system COM due to star/disk interactions is as large as 0.25 AU during the burst and averages 0.09 AU during the asymptotic phase in our simulation with the indirect potential. Superficially, this seems strikingly at odds with results reported by \citet{rice2003a}, who found a maximum radial excursion by the central star of $\sim 5 \times 10^{-3}$ AU . 

While it is true that the numerical methods are fairly different, in that we use a grid-based code and \citeauthor{rice2003a} use SPH, most of the discrepancy can probably be attributed to several other key differences: 1.) {\sl Cooling time.} \citeauthor{rice2003a} use a local cooling time tied to the local orbital speed. Specifically, they present simulations for $t_{\mathrm{cool}} = 5\Omega^{-1}$ and $t_{\mathrm{cool}} = 10\Omega^{-1}$ or, equivalently, $t_{\mathrm{cool}}/P_{\mathrm{rot}} =$ about 0.8 and 1.6, respectively. The \citeauthor{rice2003a} prescription has the effect of shortening the cooling time with decreasing radius, whereas the cooling time in our simulation is constant across the disk and is chosen to be twice the $P_{\mathrm{rot}}$ at 33 AU. 2.) {\sl Disk Structure.} Although the \citeauthor{rice2003a} disk has a similar disk-to-star mass ratio, 0.1 compared with our 0.14, their inner disk has a much higher fraction of the mass, because their disk surface density distribution is much steeper, roughly $\Sigma \sim \varpi^{-7/4}$ compared with our $\Sigma \sim \varpi^{-1/2}$. 3.) {\sl Stellar Accretion.} While very little mass moves across the cylindrical inner boundary of our grid, the SPH calculation allows its particles to approach the star, and they are incorporated into it once they enter its smoothing radius.

Because of the short cooling times in the innermost parts of the \citeauthor{rice2003a} disk simulations, the inner disks are as violently GI unstable as the outer disks. As discussed by \citet{mejia2005} in comparisons with \citet{lodato2004}, major consequences of using a constant $t_{{\mathrm cool}}/P_{{\mathrm rot}}$ prescription rather than a constant $t_{{\mathrm cool}}$ prescription in a moderate mass disk are that there is no burst, because the whole disk goes unstable at once, and that the GIs are not dominated by a few individual global modes. As our analysis has shown, it is the interaction of the star with a few discrete modes in the outer disk in both the burst and the asymptotic phases that causes the large displacements. Also, in \citeauthor{rice2003a}, the disks experience a large amount of mass accretion. In fact, the inner 2 AU of their disks is rapidly accreted onto the central star, and they assume that accreted angular momentum goes into rotation, not orbital motion. In our simulation, the inner disk remains stable, and there is essentially no accretion. The outcome of accretion onto the star, if it occurs in a simulation, could vary significantly depending on assumptions about how to treat the linear and angular momentum of any material added to the star in an asymmetrical manner. 

As with many other aspects of GI behavior \citep[e.g.,][]{pickett1998,pickett2000,pickett2003,mejia2005,boley2006}, displacement of the star is evidently quite sensitive to assumptions about disk structure and, especially, disk thermodynamics. What our indirect simulation shows is that, under conditions where GIs initiate in a burst and tend to be dominated by a few discrete modes in the outer disk, the stellar excursions about the COM can be quite large. In our simulation, the periods of the induced stellar orbital motion correspond to the CRs of the dominant modes, or about 130 and 170 years during the asymptotic and burst phases, respectively. The amplitudes of the displacements correspond to disk asymmetries with masses on the order of several to many Jupiter masses. In the asymptotic phase of our simulation, the astrometric signature would be on the order of 16 $\mathrm{\mu as/yr}$ if the system were at a distance of 100 pc, and the maximum radial velocity would be about 30 m/sec. Stellar motion of this type could be difficult to distinguish from that caused by one or more super-Jupiter protoplanets orbiting at 10's AU. 

Our simulation with the indirect potential demonstrates that discrete global GI spiral modes can, in principle, drive large stellar motion. However, whether this effect will be manifest in real disks may depend in part on whether accretion through the inner disk onto the star happens in an asymmetric fashion that counteracts the GI torques or whether stellar orbital angular momentum can be effectively removed by a jet or wind. Orbital motion of the star and of the innermost disk due to GIs might cause a precessing or helical jet/wind structure. These would be interesting questions for future research.

\section{Conclusions}
\label{sec:conclude}

The two major goals of this paper were: 1) to determine how much a proper treatment of stellar motion affects qualitative and quantitative conclusions made in previous studies by our group and 2) to decide whether stellar motion due to star/disk interaction could lead to observable consequences. We find that, although the simulations do show a weakening of disk self-torques and nonaxisymmetric structure by at most about tens of percent, simulations with and without proper treatment stellar motion are overall very similar. They exhibit the same evolutionary phases, and they both settle into similar quasi-steady asymptotic states where cooling is balanced by GI heating. The main qualitative difference is a shift of dominance from an even mode (four-armed) to an odd mode (five-armed) during the burst phase when the star's motion is properly treated. This would only matter in a case where the initial conditions at the time of onset for GIs were well understood for a particular application. We conclude that IUHG studies using a fixed star \citep[e.g.,][]{pickett2003,mejia2005,cai2006,boley2006,boley2007,boley2008,cai2008} have produced valid and reliable results regarding the overall behavior and evolution of GI-active disks.

When we account for stellar motion through an indirect potential, the motion of the star appears to be due to the nonlinear interaction of discrete odd and even global spiral modes, both during the burst and in the asymptotic phase. These interactions involve substantial disk mass at radii of about 30 AU (burst) and 24 AU (asymptotic phase) with the result that both astrometric and radial velocity disturbances are fairly large. For the asymptotic phase, we estimate about $16 ~\mathrm{\mu as/yr}$ (at 100 pc) and 30 m/s, respectively. The time signatures can involve multiple periodicities, but it would be difficult to distinguish this from a stellar response to the presence of multiple massive planets or protoplanets. Our results are in contrast to the much more modest stellar motions reported by \citet{rice2003a}, under conditions where the cooling times decrease inward in the disk, there is no burst, GIs are not dominated by a few discrete global modes, and there is significant accretion from the inner disk. This means that stellar motion induced by GIs is sensitive to various aspects of disk structure and physics. Further study is clearly warranted, especially if outbursts of GIs are common during the early phases of disk formation when disks are large and massive \citep[e.g.][]{laughlin1994,vorobyov2006,stamatellos2009,boley2009}.

\section*{Acknowledgments}

S.M. would like to thank NASA for the generous funding of the NASA Earth and Space Science Fellowship NNX07AU82H. This research was also supported in part by NASA Origins of Solar Systems grants NNG05GN11G and NNX08AK36G. The simulations and analysis herein were conducted on hardware generously provided by Indiana University Information Technology Services and was supported by the National Science Foundation under Grant No. ACI-0338618l, OCI-0451237, OCI-0535258, CNS-0521433, and OCI-0504075. This work was also supported in part by Shared University Research grants from IBM, Inc. to Indiana University. 

\bibliography{general}

\label{lastpage}

\end{document}